\documentclass[12pt,epsfig]{article}

\usepackage{mathrsfs}
\usepackage[dvips]{graphicx}
\usepackage{epsfig}
\usepackage{color}

\begin{document}

\begin{center}

{\Large STATES THAT ARE FAR FROM}

\

\

{\Large BEING STABILIZER STATES}

\vspace{10mm}

{\large David Andersson}*

%

\

{\large Ingemar Bengtsson}*

\

{\large Kate Blanchfield}*

\

{\large Hoan Bui Dang}**

\

*{\sl Fysikum, Stockholms Universitet, 106 91 Stockholm, Sweden}

%

**{\sl Perimeter Institute for Theoretical Physics and University of Waterloo,  Waterloo, Ontario, Canada}

\vspace{10mm} 

\underline{Abstract}:

\end{center}

\

\noindent Stabilizer states are eigenvectors of maximal commuting sets of 
operators in a finite Heisenberg group. States that are far 
from being stabilizer states include magic states in quantum computation, 
MUB-balanced states, and SIC vectors. In prime dimensions the latter two 
fall under the umbrella 
of Minimum Uncertainty States (MUS) in the sense of Wootters and Sussman. We 
study the correlation between two ways in which the notion of ``far 
from being a stabilizer state'' can be quantified, and give detailed 
results for low dimensions. In dimension 7 we identify 
the MUB-balanced states as being antipodal to the SIC vectors within the 
set of MUS, in a sense that we make definite. 
In dimension 4 we show that the states that come closest to being MUS with 
respect to all the six stabilizer MUBs are the fiducial vectors for Alltop 
MUBs. 

\newpage

{\bf 1. Introduction}

\

\noindent At the outset all vectors in a given Hilbert space are on the 
same footing, but in physics it frequently happens that a particular group 
of transformations is singled out for attention. Then the vectors are 
distinguished from each other by what the group does to them. In this paper 
we will look at a particularly interesting choice of the group---namely, 
finite Heisenberg groups. Once this choice is made an entire fauna of 
states springs into existence: stabilizer states, stabilizer MUBs, magic states, 
Alltop MUBs, MUB-balanced states, SICs, and more. Our concern is to see how 
they relate to each other. 

A stabilizer state, by definition, is an eigenvector of a maximal abelian 
subgroup of a finite Heisenberg group. In a certain sense---very much so in some yet-to-be-built quantum computers---stabilizer states play the role of ``classical'' 
states \cite{Gross, Emerson}.  

We let the acronym MUB stand for a complete set of $d+1$ mutually unbiased bases in dimension 
$d$. To see how they arise, note that finite Heisenberg groups are represented in $d$ 
dimensions by $d^2$ operators (not counting phase factors), and that these 
operators provide unitary 
operator bases in any dimension $d$. When a unitary operator basis can be split 
into $d+1$ disjoint sets of $d-1$ commuting operators (the unit element apart) we 
say that it forms a flower. The disjoint sets of commuting operators 
are its petals. In prime dimensions the Weyl-Heisenberg group 
defines a unique flower. In prime power dimensions the multipartite 
Heisenberg group can be displayed as a flower in more than one way. 
There are no flowers if the dimension is not a prime power, for any version 
of the Heisenberg group. When it exists, a flower provides a powerful organizing 
principle for Hilbert space. In fact the eigenbases of its petals form a MUB 
\cite{Vatan}. Since the individual vectors are stabilizer states we refer to 
such a MUB as a stabilizer MUB.

Magic states have an operational definition in universal fault-tolerant 
quantum computation \cite{Bravyi}. We do not give it here, we just observe that, 
in prime dimensions, one interesting class of magic states \cite{Campbell, Howard} 
is provided by vectors in the Alltop MUBs \cite{Alltop, Klappenecker, Kate}. 
Such vectors belong to an orbit of the Weyl-Heisenberg group forming $d$ MU bases,  
all of which are unbiased to one of the bases in the stabilizer MUB. We refer to 
the vectors in such an orbit as Alltop vectors. The Alltop MUBs 
are unitarily equivalent to stabilizer MUBs, but from the point of view of 
the Heisenberg group the Alltop vectors are very different from the stabilizer 
vectors. The latter are ``classical'', the former are ``non-classical''.  

A SIC is a set of $d^2$ unit vectors such that the absolute values squared of 
the mutual scalar products are always equal to $1/(d+1)$. 
That is, it is an equiangular tight frame \cite{Zauner, Renes}. This 
definition turns out to be very deep mathematically, and the existence of SIC 
vectors in all dimensions is conjectural only \cite{Appleby, Scott}. A few 
exceptions apart it seems that SICs always arise as orbits of the Weyl-Heisenberg 
group. This is known for a fact in dimensions 2 and 3 \cite{Hughston}. If $d=2$ 
the SIC vectors also serve as magic states \cite{Bravyi}. Whatever the dimension 
the SIC vectors are in a definite 
sense at the other end of the spectrum from the stabilizer states (for which 
each individual basis in a MUB is an orbit under the Weyl-Heisenberg group). 

To define the MUB-balanced states we need a few equations. Given $d+1$ orthonormal bases 
$\{|e_i^{(z)}\rangle \}_{i=0}^{d-1}$, we define 
the $d+1$ probability vectors $\vec{p}_{(z)}$ with 
components determined by the state vector $|\psi \rangle$ through

\begin{equation} p_{i,(z)} = |\langle e_i^{(z)}|\psi\rangle |^2 \ . \label{1} 
\end{equation} 

\noindent Here $0 \leq z \leq d$ labels the $d+1$ bases in a 
MUB, and $i$ labels the individual elements in such a basis. One can prove 
that \cite{Larsen}

\begin{equation} \sum_{z=0}^d\sum_{i=0}^{d-1}p_{i,(z)}^2 = 2 \ . 
\label{Pythagoras} \end{equation}

\noindent Wootters and Sussman \cite{Sussman} defined a Minimum Uncertainty State 
(MUS) as one for which 

\begin{equation} \sum_{i=0}^{d-1}p_{i,(z)}^2 = \frac{2}{d+1} \ , \label{ApDaFu}
\end{equation}

\noindent for every MU basis individually. Now consider a state as a vector in 
Bloch space, the set of unit trace Hermitean matrices with its origin at the 
maximally mixed state. A probability vector arises, as in eq. (\ref{1}), 
from an orthogonal projection of a Bloch vector to a plane spanned by the states in 
a Hilbert space basis. Then eq. (\ref{Pythagoras}) 
follows from Pythagoras' theorem and eqs. (\ref{ApDaFu}) mean that the 
length of the Bloch vector when projected orthogonally onto the plane defined by 
a MU basis is independent of which basis we pick \cite{DMA}. It is understood 
that the MUB with respect to which the MUS is defined is a stabilizer MUB, and 
then we see that there is a sense---at 
least in prime dimensions, where the stabilizer MUB is unique---in which 
a MUS is indeed far from 
being a stabilizer state. In a Hilbert space of dimension $d$, one expects the 
set of MUS to form a continuous set of real dimension $d-2$, while the set 
of all pure states has real dimension $2d-2$. 

A MUB-balanced state is defined as one for which the $d+1$ probability vectors 
$\vec{p}_z$ are identical up to permutations of their components \cite{ASSW}. 
Such states form a 
distinguished discrete set inside the continuous set of MUS. They can arise as 
eigenvectors of a MUB cycling operators, operators that cycle through all the $d+1$ 
bases in a MUB.  
MUB-balanced states are known to exist if $d = 2^n$ where unitary MUB-cyclers exist 
\cite{Sussman, Seyfarth}, and if $d = (\mbox{prime})^n = 3$ modulo 4 where 
anti-unitary MUB-cyclers exist \cite{ASSW, ABD}. These states have some intriguing 
and useful properties \cite{ASSW, ABD, Anton}. For odd $d$ the parity of these states 
(in the language used in 
connection with discrete Wigner functions, say) is opposite to that of the 
stabilizer states, and there is a sense in which this property alone makes 
them far from being stabilizer states: they maximally violate a certain 
non-contextuality inequality used to demarcate states that behave 
``classically'' in quantum computing \cite{Joe}.

Interestingly, if they exist, and if the dimension of Hilbert space is a prime 
number, the SIC vectors 
form another distinguished discrete subset of the set of MUS \cite{ADF}. In fact 
this was the first concrete hint (out of three \cite{Huangjun, Mark}) of a 
connection between MUBs and SICs in prime dimensions larger than 3. To 
complete the story we mention that another interesting notion of ``far from 
being a stabilizer state'' is that of 
maximal mana states \cite{Veitch}. These are SIC vectors if $d = 2,3$ but if 
$d = 5$ they are not even MUS \cite{Gelo}. They will not be discussed here. 

In the next section we introduce quantitative measures of how far a 
given state is from being a MUS, and how far it is from being a Weyl-Heisenberg 
covariant SIC. We also derive an inequality, relating these measures in 
prime dimensions. Then we study the correlation between them, 
and see where MUB-balanced states and Alltop vectors fit into the 
resulting picture. This is mostly done numerically, beginning with the prime 
dimensions $d =$ 2, 3, 5, 7. We make an aside on ``Zauner subspaces'' (of 
interest in the SIC existence problem \cite{Zauner}). Finally we come to 
$d =$ 4, where the 
question of the correlation between the measure of MUSness relative to two different 
stabilizer MUBs arises. By the time we formulate our conclusions, in section 
7, the beginnings of an orderly pattern for the states we discuss will have emerged. 

\newpage

{\bf 2. Quantitative measures, and an inequality}

\

\noindent In any dimension $d$ the Weyl-Heisenberg group $H(d)$ is generated 
by two elements $X$ and $Z$ which in themselves generate cyclic subgroups 
of order $d$, and obey 

\begin{equation} ZX = \omega XZ \ , \end{equation}

\noindent where $\omega = e^{2\pi i/d}$. Once the generator $Z$ is given in 
diagonal form the unitary representation is unique. One finds that 

\begin{equation} Z|e_i\rangle = \omega^i|e_i\rangle \ , \hspace{8mm} 
X|e_i\rangle = |e_{i+1} \rangle \ , \end{equation}

\noindent where integers modulo $d$ are used to label the states. In odd prime 
dimensions it is convenient to work with the $d^2$ displacement operators 

\begin{equation} D_{ij} = \omega^{\frac{1}{2}ij}X^iZ^j \ , \end{equation}

\noindent where $1/2$ is the multiplicative inverse of 2 modulo $d$. The 
full story can be found in many places, say in ref. \cite{Appleby}. 

We define

\begin{equation} f_{\rm SIC}(\psi ) = 
\sum_{(i,j)\neq (0,0)}\left( |\langle \psi|D_{ij}|\psi\rangle 
|^2- \frac{1}{d+1}\right)^2 \ . \label{fSIC} \end{equation}

\noindent This octic expression in the components of the unit vector $|\psi\rangle $ 
is also known as a frame potential \cite{Fickus}, and is familiar from the study 
of 2-designs, except that we rescaled and then shifted it 
so that $f_{\rm SIC} = 0$ if and only if $|\psi\rangle$ is a SIC vector. The SIC 
itself is obtained by acting on such a vector with all the displacement operators---the 
absolute values squared of all the scalar products equal $1/(d+1)$ in this case.

Given a MUB, with bases labelled by $z$, we also define 

\begin{equation} f_{\rm MUS}(\psi ) 
= \sum_{z=0}^d \left( \sum_{r=0}^{d-1}|\langle e_r^{(z)}|\psi\rangle 
|^4 - \frac{2}{d+1}\right)^2 \ . \label{fMUS} \end{equation}

\noindent This function is again an octic polynomial, and it vanishes if and 
only if $|\psi \rangle$ is a MUS, as defined in the introduction. 

In prime dimensions these two measures are related by the inequality 

\begin{equation} f_{\rm SIC} \geq \frac{d^2}{d-1}f_{\rm MUS}  \ , \label{David} \end{equation} 

\noindent with equality for all states if and only if $d = 2,3$. 
For $d = 2$ it is easy to show by means of an explicit calculation that equality holds. 
Now let $d$ be an odd prime. The idea of the proof is to split $f_{\rm SIC}$ into $d+1$ 
terms, one for each petal. Focus on one such maximal abelian subgroup, denote 
its generator by $Z$, and diagonalize. In this basis $|\psi\rangle $ is 

\begin{equation} \psi = \left( \begin{array}{c} \sqrt{p_0} \\ \sqrt{p_1}e^{i\mu_1} \\ 
\vdots \\ \sqrt{p_{d-1}}e^{i\mu_{d-1}} \end{array} \right) \ , \hspace{10mm} 
p_i \geq 0 \ , \hspace{5mm} \sum_{i=0}^{d-1}p_i = 1 \ . \label{pure} \end{equation}

\noindent The calculation is exactly the same in all the $d+1$ eigenbases. Therefore 
it will be enough to show that 

\begin{equation} \sum_{j=1}^{d-1}\left( |\langle \psi|Z^j|\psi \rangle |^2 - 
\frac{1}{d+1}\right)^2 \geq \frac{d^2}{d-1}
\left( \sum_{r=0}^{d-1} p_r^2 - \frac{2}{d+1}\right)^2 \ . \label{olikhet} \end{equation}  

\noindent Using the standard representation of $Z$ we observe that 

\begin{equation} |\langle \psi|Z^j|\psi \rangle |^2 = \sum_{k=0}^{d-1}p_k^2 + 
\sum_{k\neq l}\omega^{j(k-l)}p_kp_l = \sum_{k=0}^{d-1}p_k^2 
+ \sum_{k=1}^{\frac{d-1}{2}}(\omega^{jk} + \omega^{-jk}) \Delta_k \ , \end{equation}

\noindent where 

\begin{equation} \Delta_k = \sum_{m = 0}^{d-1}p_mp_{m+k} \ . \end{equation}

\noindent After an amount of manipulation we find that the inequality 
(\ref{olikhet}) holds if and only if 

\begin{eqnarray} (d-1)\sum_{k=1}^{\frac{d-1}{2}}\Delta_k^2 \geq \frac{1}{2} \left( 1 - 
\sum_{k=0}^{d-1}p_k^2\right)^2 = 2\left( \sum_{k=1}^{\frac{d-1}{2}}\Delta_k\right)^2 
\nonumber \\ \Leftrightarrow \nonumber \\
\sum_{k = 1}^{\frac{d-1}{2}}\sum_{l = 1}^{\frac{d-1}{2}}(\Delta_k^2 
+ \Delta_l^2) \geq 2\sum_{k = 1}^{\frac{d-1}{2}}\sum_{l = 1}^{\frac{d-1}{2}}\Delta_k\Delta_l 
\hspace{12mm} \\ \Leftrightarrow \nonumber \\
\sum_{k=1}^{\frac{d-1}{2}}\sum_{l=1}^{\frac{d-1}{2}}
(\Delta_k-\Delta_l)^2 \geq 0 \ , \hspace{18mm} \nonumber \end{eqnarray}

\noindent and we are done. It is clear that equality holds for all probability vectors 
if and only if $d = 3$. In higher dimensions more than one distinct $\Delta_k$ may 
occur.

States that saturate the inequality have an interesting geometrical property. Let 
such a state serve as the fiducial state in a Weyl-Heisenberg orbit. In Bloch 
space, project the $d^2$ (pure) density matrices in the orbit orthogonally onto the plane 
spanned by an eigenbasis of a cyclic subgroup, 
that is to a plane defined by a MU basis. Denote the generator of this cyclic 
subgroup with $Z$. When this generator acts on a state it does not affect its 
image under the projection. Thus only $d$ distinct points will appear 
when we project the entire orbit. Denote the complementary generator 
with $X$. Its effect on the projection is to permute the entries of the probability 
vector cyclically, $p_i \rightarrow p_{i-1}$. The state saturates the inequality 
if and only if $\Delta_k$ takes the same value for all $k$, for all eigenbases. 
But the $\Delta_k$ 
are precisely the mutual scalar products of the probability vectors. Hence the 
$d$ projections of the orbit sit at the vertices of a regular 
simplex for such a fiducial state. The same argument applies to all the $d+1$ eigenbases.

\begin{figure}[h]
        \centerline{ \hbox{
                \epsfig{figure=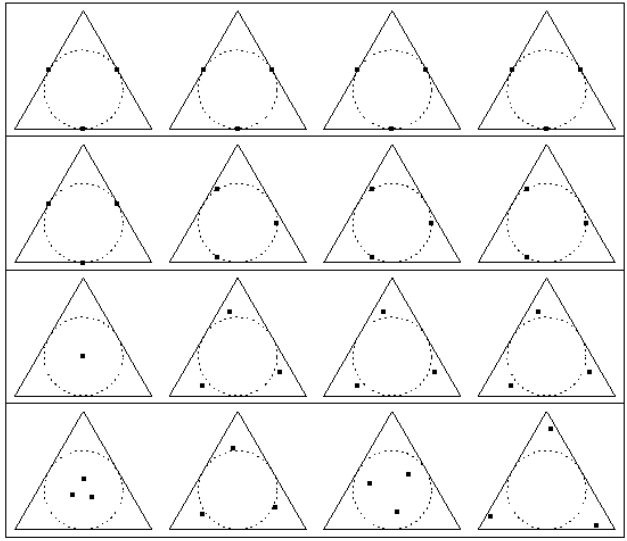,width=85mm}}}
        \caption{\small For $d = 3$ we show the orthogonal projections of the 9 
        vectors in a Weyl-Heisenberg orbit onto the four MUB simplices. It is always 
        the case that three images coincide, so we see only 3 points in each projection. 
        From top 
        to bottom we see a MUB-balanced SIC, a generic SIC, an Alltop orbit, and 
        an orbit whose fiducial vector is a random state.}
        \label{fig:grafik}
\end{figure}

SIC vectors do saturate the inequality for all prime dimensions, and being MUS they have 
the additional property that all the simplices we see in the $d+1$ projections 
are of the same size \cite{DMA, ADF}. Alltop vectors also saturate the inequality 
as one can see from a formula given by Khaterinejad \cite{Mahdad}, and they have 
the additional property that $d$ out of $d+1$ projected simplices share the same 
size and the same orientation, as follows from the fact that every Alltop vector is 
left invariant by a unitary operator that cycles through $d$ of the bases in 
the stabilizer MUB \cite{Kate}. 
A glance at Fig. \ref{fig:grafik} may clarify this description. 

\begin{table}
\caption{Some states that saturate the inequality ($d$ is prime)}
  \smallskip
\hskip 1.6cm
{\renewcommand{\arraystretch}{1.67}
\begin{tabular}
{|l| c|c|c|}\hline \hline
States & $f_{\rm MUS}$  & $f_{\rm SIC}$  & Remark \\
 \hline
Stabilizer states & $\frac{(d-1)^2}{d(d+1)}$
         & $\frac{d(d-1)}{d+1}$ & classical states \\
Alltop vectors & $\frac{(d-1)^2}{d^3(d+1)}$
         & $\frac{d-1}{d(d+1)}$ & magic states \\
SIC vectors & $0$
          & $0$ & mysterious states \\
\hline \hline
\end{tabular}
}
\label{tab:states}
\end{table} 

Table \ref{tab:states} gives some precise information. For comparison it is also 
useful to know that the Fubini-Study averages over all Hilbert space are \cite{Helena}

\begin{equation} \left< f_{\rm SIC}\right>_{\rm FS} = \left\{ \begin{array}{l} 
\frac{d(d-1)}{(d+2)(d+1)} \ \mbox{if} \ d \ \mbox{is odd} \\ \\ 
\frac{d^2}{(d+3)(d+1)} \ \mbox{if} \ d \ \mbox{is even} \end{array} \right. \label{FS1} \end{equation} 

\begin{equation} \left< f_{\rm MUS}\right>_{\rm FS} = \frac{4(d-1)}{(d+3)(d+2)(d+1)} \ . 
\label{FS2} \end{equation}

\noindent The latter tends to zero with growing dimension, as is reasonable.

\

\

{\bf 3. Dimensions 2 and 3}

\

\noindent From the previous section it is clear that in dimensions 2 and 3 
every Minimum Uncertainty State is a SIC vector (and the converse holds  
\cite{FS,Hughston}). In dimension 2 it is also true 
that each of the eight MUS is MUB-balanced. 

In 3 dimensions it is known that every Weyl-Heisenberg covariant SIC is 
equivalent (to be precise, equivalent under the extended Clifford group \cite{Appleby}) 
to one obtained from a fiducial vector with components \cite{Zauner}

\begin{equation} \psi (\sigma) = \left( \begin{array}{c} 0 \\ 1 \\ - e^{i\sigma} 
\end{array} \right) \ , \end{equation}

\noindent where $\sigma \in [0, 2\pi /6]$ . 
This is a MUB-balanced state if and only if $\sigma = 0$, and incidentally a 
maximal mana state if $\sigma = 0$ or $2\pi/6$ \cite{Veitch}. Since all states 
in these dimensions saturate the inequality (\ref{David}) we see regular triangles 
in all projections of every Weyl-Heisenberg orbit onto the MU planes, but they 
have the same size if and only if the fiducial vector is a SIC. See Fig. \ref{fig:grafik}. 

\

\

{\bf 4. Dimensions 5 and 7}

\

\begin{figure}[t]
        \centerline{ \hbox{
                \epsfig{figure=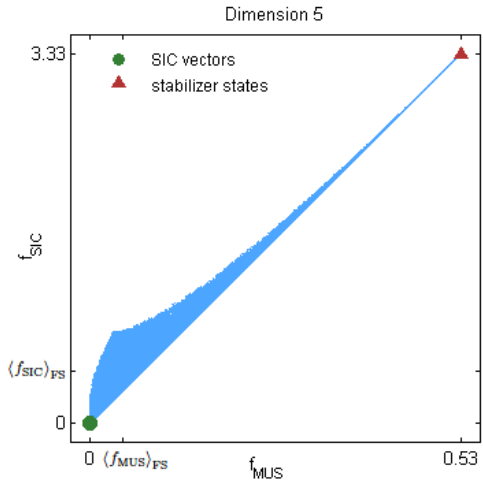,width=70mm}}}
        \caption{\small The correlation between SICness and MUSness in dimension 5. 
        The plot uses $6\cdot 10^5$ vectors, chosen in equal numbers at random, close 
        to a stabilizer vector, and close to a SIC vector.}
        \label{fig:sicmus5}
\end{figure}

\begin{figure}[ht]
        \centerline{ \hbox{
                \epsfig{figure=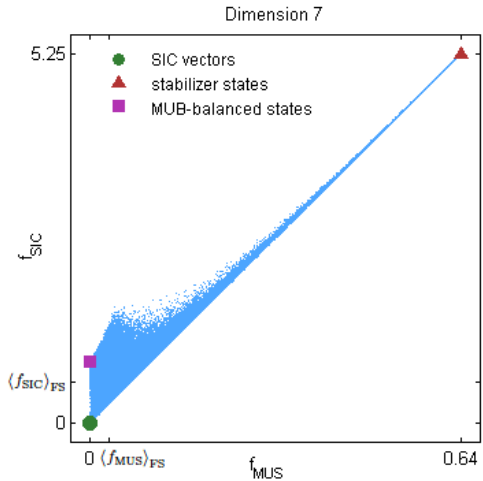,width=70mm}}}
        \caption{\small The correlation between SICness and MUSness in dimension 7. 
        The plot uses $8\cdot 10^5$ vectors, chosen in equal numbers at random, close 
        to a stabilizer vector, close to a SIC vector, and close to a MUB-balanced state.}
        \label{fig:sicmus7}
\end{figure}

\noindent In dimensions 5 and 7 we begin by choosing vectors at 
random, computing the values that $f_{\rm SIC}$ and $f_{\rm MUS}$ take for 
them, and plotting the results. See Figs. \ref{fig:sicmus5} and \ref{fig:sicmus7}. 
The results look 
rather similar for these two dimensions (and we have checked that they continue 
to look similar for $d =$ 11, 13, 17, and 19, although our coverage in higher 
dimensions is not very good). In particular, all the points fall inside a region whose 
boundary consists of four segments. One of them is the simplex line, where 
the inequality (\ref{David}) is saturated. It gets its name from the 
geometric interpretation in section 2. It begins at the origin (where the 
SIC vectors sit) and ends at the stabilizer states. 
There it joins a curved segment, and we have convinced ourselves that the 
states that end up on this segment consist of superpositions of two orthogonal 
stabilizer states. This segment ends at a point where the superposition 
is of equal weight, and is then joined by another segment that we have 
not understood. Finally Minimum Uncertainty States form the segment that 
lies on the vertical axis. 

One obvious distinguished point in the plots is given by Minimum Uncertainty 
States that maximize $f_{\rm SIC}$, 
given that $f_{\rm MUS} = 0$. We identified this point using the NMaximize 
routine in Mathematica. When $d = 5$ we find states that we do not recognize. 
However, when $d = 7$ we do recognize them. They are MUB-balanced states. 
Such states exist when $d$ is an even prime power \cite{Sussman} or 
an odd prime power equal to 3 modulo 4 \cite{ASSW, ABD}, but at least conjecturally 
not otherwise. We find it somewhat 
remarkable that these states are ``antipodal'' to the SIC vectors within the set 
of Minimum Uncertainty States. 

To be precise, the numerical calculation shows that, when $d = 7$, 

\begin{equation} f_{\rm MUS} = 0 \hspace{5mm} \Rightarrow \hspace{5mm} 0 \leq 
f_{\rm SIC} \leq \frac{7}{8} \ . \end{equation}

\noindent We do not have an analytical proof of this statement, and we do not have 
a proof that all MUS with 
$f_{\rm SIC} = 7/8$ are indeed MUB-balanced, but we generated 26 such 
states numerically by maximizing $f_{\rm SIC}$ under the constraint $f_{\rm MUS} = 0$, 
and the statement was true in all cases. This result does not seem to generalize 
to any higher dimension. In dimension 11 we 
numerically generated one vector maximizing $f_{\rm SIC}$ under the 
constraint $f_{\rm MUS} = 0$. The resulting vector obeys 
the constraint to a precision of $10^{-20}$, but it has a value of $f_{\rm SIC}$ 
which is more than twice as large as that attained by the MUB-balanced states 
in this dimension. With less precision we also generated a MUS state in dimension 
19 whose $f_{\rm SIC}$ value exceeds that of the MUB-balanced states there. 

\begin{figure}
   \centerline{ \hbox{ 
        \epsfig{figure=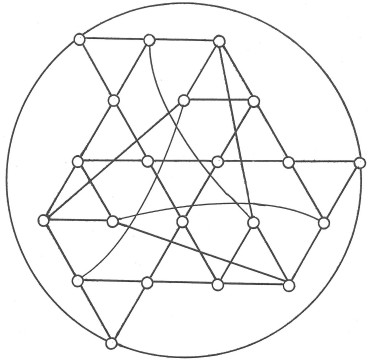, width=50mm}}}
    \caption{\small Orthogonality graph for the 21 MUB balanced states in a 
    negative parity eigenspace in dimension 7. Technically this is a vertex 
    transitive and perfect graph.}
   \label{fig:MUBbalans}
\end{figure}

As mentioned in the introduction, MUB-balanced states in odd prime dimensions $d$ 
have negative parity. There are $d^2$ negative parity eigenspaces 
altogether, related by the Weyl-Heisenberg group, and each negative parity 
eigenspace contains $d(d-1)/2$ MUB-balanced states \cite{ABD}. 
The orthogonality relations between 
them form interesting patterns, shown for the 21 vectors we can find in $d = 7$ in 
Fig. \ref{fig:MUBbalans}. 

We find it interesting that we encountered MUB-balanced states in dimension 7 but 
not in dimension 5, and that the states we did find in dimension 7 were among 
those already known (which can be constructed as eigenvectors of MUB-cycling 
anti-unitaries belonging to the extended Clifford group \cite{ASSW, ABD}). We 
regard this as circumstantial evidence for 
the conjecture that all states of this kind have been identified already. 

\

\

{\bf 5. A look at Zauner subspaces}

\

\noindent This section is an aside partly motivated by an unexplained property 
shared by every Weyl-Heisenberg covariant SIC so far constructed \cite{Scott}, 
namely that---as conjectured by Zauner \cite{Zauner}---every 
vector in such an orbit is left invariant by an element of the unitary 
automorphism group of the Weyl-Heisenberg group, having order 3. Thus they sit in 
special subspaces, known as Zauner subspaces. This section is 
also motivated by a naive question: what do the level surfaces of the function 
$f_{\rm SIC}$ actually look like? 

Unfortunately the dimension of 
the Hilbert space is typically too large for visualization. However, if the 
Hilbert space has dimensions 4 or 5 the Zauner subspace has dimension 2 only, 
and this can be visualized as a Bloch sphere. The resulting pictures of $f_{\rm SIC}$ 
are quite complex \cite{David}, which begins to explain 
why finding SIC vectors using numerical methods is a difficult art \cite{Scott}. When 
$d = 5$ the one-parameter family of 
Minimum Uncertainty States in the Zauner subspace can be solved for exactly \cite{David}.

\begin{figure}[h]
        \centerline{ \hbox{
                \epsfig{figure=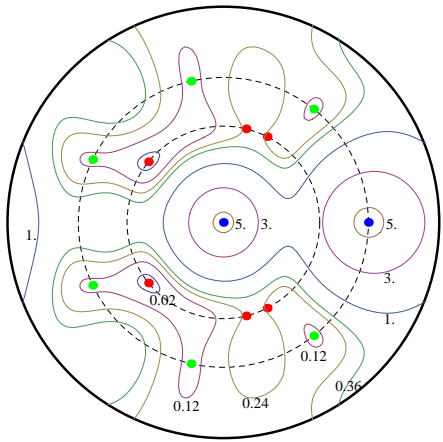,width=60mm}}}
        \caption{\small Stereographic projection of a hemisphere, 
        or equivalently of the real subspace of the Zauner subspace, in dimension 7. 
        The positions of 6 Alltop vectors (at latitude $22^\circ$) and 6 MUS states 
        (at latitude $42^\circ$) are shown against a background of contour curves for                   $f_{\rm SIC}$. Maxima ($f_{\rm SIC} = 5.24$) occur at 
        two eigenvectors of the Weyl-Heisenberg group (one of them sits at the 
        pole), there are several local minima, 
        and nothing special happens at the MUS states unless they are also SIC vectors 
        (as happens for two out of six).}
        \label{fig:David2}
\end{figure}

In Hilbert space of dimension 7 the Zauner subspace is three dimensional but contains a 
real subspace of considerable interest, and this real subspace defines a two 
dimensional real projective space. Fig. \ref{fig:David2} is a map of the real 
Zauner subspace in dimension 7, 
making use of the fact that real projective 2-space can be viewed 
as a sphere with antipodal points identified, or equivalently as 
the upper hemisphere of a sphere. Stereographic coordinates are used. By solving the polynomial equations that define Minimum Uncertainty 
States we have verified that there are exactly 6 such states within this real subspace. They 
are marked on the map. Only two of them are SIC vectors \cite{Appleby}. 
Interestingly this subspace also contains 6 Alltop vectors, with a specific value 
of $f_{\rm SIC}$, and their position are 
given as well. In fact Alltop vectors will occur in the analogous subspace in all 
prime dimensions $d = 1$ modulo 3 \cite{Mark}, while SIC vectors occur there 
only in a few cases \cite{Mahdad}.

\

\

{\bf 6. Dimension 4}

\

\begin{figure}[h]
        \centerline{ \hbox{
                \epsfig{figure=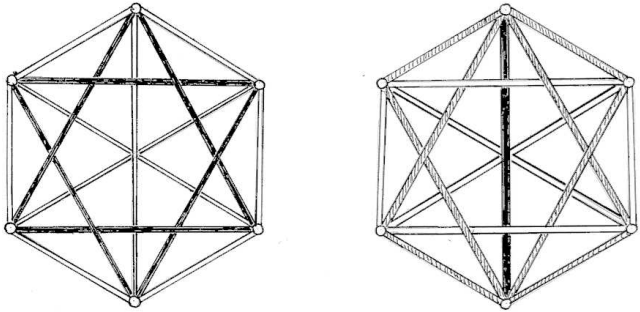,width=85mm}}}
        \caption{\small The 15 edges of the complete 6-graph are the bases defined by 
        altogether 15 maximal abelian subgroups. The vertices are MUBs. 
        The 5 edges meeting at a vertex represent the bases in that MUB. 
        On the left: In the computational basis 6 bases are maximally entangled, and 
        they are drawn in black. On the right: the computational basis is drawn in 
        black, 8 bases form Hadamard matrices, and 6 form 
        sparse matrices drawn in white.}
        \label{fig:Israel}
\end{figure}

\noindent Since 4 is a prime power we have a choice between two non-isomorphic 
Heisenberg groups when $d = 4$. The one having a SIC as an orbit is the usual 
Weyl-Heisenberg group $H(4)$, while the one underlying the mutually unbiased 
bases is the bipartite direct product $H(2)\times H(2)$. In fact the bipartite group 
admits 15 maximal abelian subgroups having altogether 60 stabilizer 
states as eigenvectors. The latter can be organized into stabilizer 
MUBs in 6 different but unitarily equivalent ways. This well known situation is 
summarized in Fig. \ref{fig:Israel}, and elsewhere \cite{Gunnar}. Since the 
SIC in this dimension is covariant under $H(4)$, and no canonical identification 
of the computational bases of the two groups is known, it plays no role in this 
paper.

Minimum Uncertainty States, and the function $f_{\rm MUS}$, can be defined 
relative to any MUB. In dimension 4 we have 6 stabilizer MUBs to choose from. 
The definition of the frame function 
$f_{\rm SIC}$ can be used in dimension 4 provided it is modified so that the 
sum in eq. (\ref{fSIC}) runs over the non-trivial elements of the bipartite 
Heisenberg group. We have checked that the inequality (\ref{David}) with 
$d = 4$ holds in this case too, and that the maximum of these functions 
is attained by the vectors in the relevant MUB. To be precise, for the six 
different cases in which $f_{\rm MUS}$ can be defined it is true that

\begin{equation} f_{\rm SIC} \geq \frac{16}{3}f_{\rm MUS} \ . \end{equation}

\noindent The bound can be saturated. Moreover there holds, in general, that  

\begin{equation} 0 < f_{\rm SIC} \leq \frac{12}{5} \ . \end{equation}

\begin{figure}[h]
        \centerline{ \hbox{
                \epsfig{figure=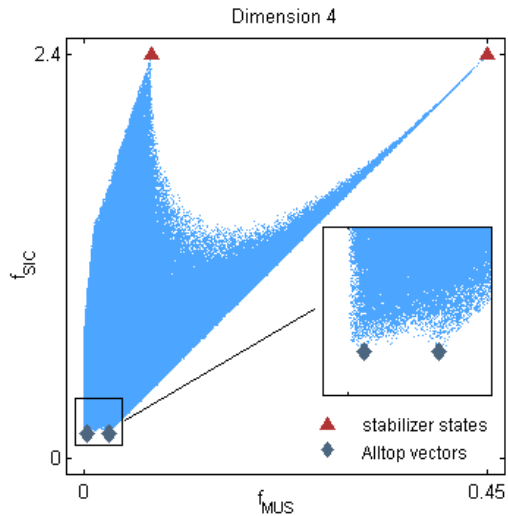,width=75mm}}}
        \caption{\small The correlation between the SICness and MUSness in 
        dimension 4. The plot uses $5\cdot10^5$ random 
        vectors. For stabilizer states it matters whether they belong to the MUB 
        with respect to which $f_{\rm MUS}$ is defined. {\it Mutatis mutandis}             this 
        applies to the Alltop vectors too.}
        \label{fig:sicmusd4}
\end{figure}

\noindent The upper bound is saturated by the stabilizer states, but this time 
the lower bound cannot be reached because the bipartite Heisenberg group 
does not admit a SIC as an orbit. A plot is shown in Fig. \ref{fig:sicmusd4}.

There are MUB-balanced states in $d = 4$, and when the MUB is a stabilizer MUB they 
can be constructed as eigenvectors 
of an element in the Clifford group of order 5 \cite{Sussman}. Such unitaries cycle through 
the bases in one of the MUBs, and they move bases in one of the other five MUBs through 
these five. Since there are six MUBs altogether we are in fact dealing with six different 
MUS-functions $f^{(i)}_{\rm MUS}$, where $i$ labels the particular MUB with respect to 
which the functions are defined. A MUB-balanced state is a MUS only with respect 
to one of them. If the state is balanced with respect to the first MUB one finds 

\begin{equation} f_{\rm SIC} = 0.32 \ , \hspace{6mm} f^{(1)}_{\rm MUS} = 0 \ , 
\hspace{6mm} f^{(2)}_{\rm MUS} = \dots = f^{(6)}_{\rm MUS} = 0.032 \ . \end{equation}

\noindent Judging from Fig. \ref{fig:sicmusd4} these are not very remarkable values.

\begin{figure}[ht]
        \centerline{ \hbox{
                \epsfig{figure=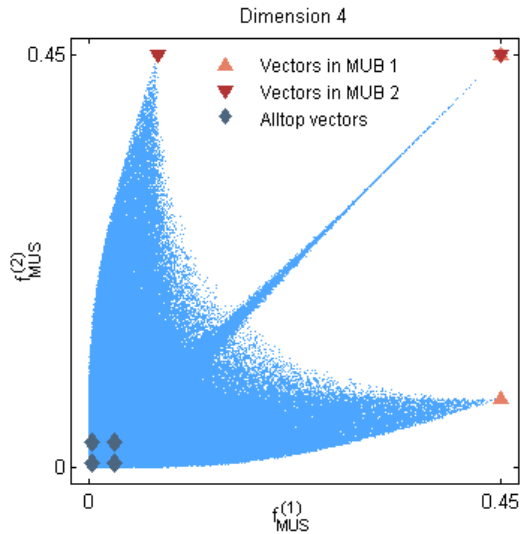,width=75mm}}}
        \caption{\small The correlation between the MUSness as defined with 
        respect to any two different MUBs. The plot uses $5\cdot10^5$ random 
        vectors.}
        \label{fig:musmus}
\end{figure}

A new question is now posing itself, because we can ask for the correlation between 
two different MUS-functions $f^{(i)}_{\rm MUS}$. This results in the 
stingray shown in Fig. \ref{fig:musmus}. 
Inspection of this plot may suggest the possibility of finding a state which 
is MUS relative to two stabilizer MUBs, but in fact no such states exist. Still 
there will exist states that minimize the sum of the six different 
functions $f_{\rm MUS}$ that we obtain from the six stabilizer MUBs. In this very sense, 
these states deserve to be regarded as being as far 
from stabilizer states as any state can be.

\begin{table}
\caption{Minimizing the MUSness with respect to more than one MUB.}
  \smallskip
\hskip 1.6cm
{\renewcommand{\arraystretch}{1.1}
\begin{tabular}{||c|c||} \hline \hline
Number of MUBs & minimum$\left( \sum f^{(i)}_{\rm MUS}\right) $ \\ \hline 
1 & 0.0000000000 \\ 
2 & 0.0041666666 \\ 
3 & 0.0102012357 \\
4 & 0.0187500000 \\ 
5 & 0.0468749999 \\ 
6 & 0.0749999999 \\ \hline 
\hline \end{tabular}}
\label{tab:two}
\end{table} 

We used Matlab to minimize, numerically, the sum of the $f_{\rm MUS}$-functions 
for between 1 and 6 stabilizer MUBs. 10 000 random initializations were used 
in each case. The results are reported in Table \ref{tab:two}. 

We find that the states that minimize the sum over 4 or more $f_{\rm MUS}$ 
have the property that if we act on them with the bipartite Heisenberg group 
the resulting orbits form sets of four mutually unbiased bases. Therefore 
these states are fiducial vectors for Alltop MUBs. Each orbit 
forms a MUB when taken together with one 
of the 15 stabilizer bases. Using an exact expression for such 
fiducials \cite{Kate} we find that 

\begin{equation} f_{\rm MUS} = 
\left\{ \begin{array}{lll} \frac{9}{320} \approx 0.0281 
& \mbox{if the Alltop and stabilizer MUBs share a basis} \\ \\ 
\frac{3}{640} \approx 0.0047 
& \mbox{otherwise.} \end{array} \right. 
\ \end{equation}

\noindent Since the stabilizer basis to which the Alltop orbit is unbiased 
occurs in two out of the four stabilizer MUBs (see Fig. \ref{fig:Israel}) it 
follows that the minimum of $f_{\rm MUS}$ summed over all six stabilizer 
MUBs is 

\begin{equation} \mbox{min} \left( \sum_{i=1}^6f^{(i)}_{\rm MUS} \right) = 
4\cdot \frac{3}{640} + 2\cdot \frac{9}{320} = \frac{3}{40} = 
0.075 \ , \end{equation}

\noindent in agreement with the table. We conclude that there is a 
definite sense in which the prize for being 
as far as possible from the stabilizer states in $d = 4$ goes to the Alltop 
vectors.

Interestingly, the Alltop vectors also minimize $f_{\rm SIC}$ as defined using 
the bipartite Heisenberg group. We have not proved that this is the case, but 
it emerges clearly from Fig. \ref{fig:sicmusd4}. To remove any doubt we performed 
a numerical minimization of $f_{\rm SIC}$. We made 1000 trials and 
ended up, each time, with vectors having the values of $f_{\rm SIC}$ and 
$f_{\rm MUS}$ that obtain for the Alltop vectors. 

It is perhaps worth noticing 
that the situation in dimension 8 must be different, since there 
does exist a SIC covariant under $H(2)\times H(2)\times H(2)$ in this dimension 
\cite{Hoggar}. But this is an exceptional case \cite{Godsil}. 

\

\

{\bf 7. Conclusions}

\

\noindent We have walked through woods inhabited by states 
classified by a Heisenberg 
group into stabilizer states, Alltop vectors, MUB-balanced states, SIC vectors, 
and more. The stabilizer states are peaceful ``classical'' states from the 
point of view of some quantum computers, while the others are ``wild'' and 
in some sense essentially quantum. 

Things are fairly simple in dimension 3 where every Minimum 
Uncertainty State is a SIC, one of which is composed of MUB-balanced states. 
In dimension 5 MUB-balanced states 
presumably do not exist, but in dimension 7 they do, and turn out to be---in a 
sense we made precise---antipodal to the SIC vectors within the set of 
Minimum Uncertainty States. We regard the 
results reported 
here as circumstantial evidence that the set of all MUB-balanced states coincides 
with the set of those that are already known \cite{Sussman, ASSW, ABD}. 

In dimension 4 an interlocking system of six MUBs 
can be constructed from the stabilizer states. We find a clear 
sense in which the Alltop vectors are the states that are as far from being stabilizer 
states, and as close to being SIC vectors under the relevant group, as any state can be.

With the caveat that most of our discussion has been confined to 
low dimensions, we believe that we have introduced a certain amount 
of order into the question of providing meaning to the expression 
``far from being a stabilizer state''. The caveat is of course an important one. 
Many things are unknown in higher dimensions, including even the existence of 
Alltop MUBs in even prime power dimensions larger than four, and the existence 
of SICs in three digit dimensions and larger.  
Moreover a truly satisfactory picture requires further study also for the 
dimensions we do study. We have not discussed maximal mana states, or more 
generally how the states we have discussed sit relative to the set of mixed 
states with positive Wigner function. This set plays a special role in quantum 
computation. We hope to address some of these issues in the future.

\

\noindent \underline{Acknowledgement}: 

\noindent We thank Ad\'an Cabello for an 
analysis of the graph given in Fig. \ref{fig:MUBbalans}, and Marcus Appleby 
for many discussions. 

\
 
{\small

\end{document}